\documentclass[aip,jcp,reprint,letter]{revtex4-1}
\usepackage{graphicx,color}
\usepackage{amsmath}
\usepackage{mathrsfs}

\draft 

\begin{document}
\newcommand{\calA}{ {\mathcal A}}
\newcommand{\calC}{ {\mathcal C}}
\newcommand{\calD}{ {\mathcal D}}
\newcommand{\calE}{ {\mathcal E}}
\newcommand{\calG}{ {\mathcal G}}
\newcommand{\calH}{ {\mathcal H}}
\newcommand{\calL}{ {\mathcal L}}
\newcommand{\calO}{ {\mathcal O}}
\newcommand{\calP}{ {\mathcal P}}
\newcommand{\calS}{ {\mathcal S}}
\newcommand{\calZ}{ {\mathcal Z}}
\newcommand{\calN}{ {\mathcal N}}

\newcommand{\rsfsL}{ {\mathscr L}}


\title{Irreversible Markov chain Monte Carlo algorithm for self-avoiding walk} 

\author{Hao Hu}

\affiliation{National Laboratory for Physical Sciences at Microscale and Department of Modern Physics, University of Science and Technology of China, Hefei, Anhui 230026, China}
\affiliation{State Key Laboratory of Theoretical Physics, Institute of Theoretical Physics, Chinese Academy of Sciences, Beijing 100190, China}
\affiliation{School of Chemical and Biomedical Engineering, Nanyang Technological University, Singapore 637459}

\author{Xiaosong Chen}
\affiliation{State Key Laboratory of Theoretical Physics, Institute of Theoretical Physics, Chinese Academy of Sciences, Beijing 100190, China}

\author{Youjin Deng}
\email{yjdeng@ustc.edu.cn}

\affiliation{National Laboratory for Physical Sciences at Microscale and Department of Modern Physics, University of Science and Technology of China, Hefei, Anhui 230026, China}
\affiliation{State Key Laboratory of Theoretical Physics, Institute of Theoretical Physics, Chinese Academy of Sciences, Beijing 100190, China}





\date{\today}

\begin{abstract}
We formulate an irreversible Markov chain Monte Carlo algorithm for the self-avoiding walk (SAW),
which violates the detailed balance condition and satisfies the balance condition. 
Its performance improves significantly compared to that of the Berretti-Sokal algorithm,
which is a variant of the Metropolis-Hastings method.
The gained efficiency increases with the spatial dimension (D), 
from approximately $10$ times in 2D to approximately $40$ times in 5D. 
We simulate the SAW on a 5D hypercubic lattice with periodic boundary conditions, 
for a system with a linear size up to $L=128$, 
and  confirm that as for the 5D Ising model, the finite-size scaling of the  SAW is governed by renormalized exponents $\nu^*=2/d$ 
and $\gamma/\nu^*=d/2$. The critical point is determined, which is approximately $8$ times more precise than the best available estimate.

\end{abstract}


\maketitle 

\section{Introduction}
\label{secIntro}

The self-avoiding walk (SAW) serves as a paradigmatic model in polymer 
physics (see e.g. Ref.[\onlinecite{deGennes79}] and references therein). 
It is equivalent to the $n \to 0$ limit of 
the O$(n)$ model \cite{deGennes72} and plays an important role in the study
of critical phenomena. 
In the grand-canonical ensemble, 
the length of a walk can fluctuate and 
the SAW model is defined by the following partition sum 
\begin{equation}
\calZ = \sum_{\omega} x^{|\omega|} \;,
\label{partition_sum}
\end{equation}
where $|\omega|$ is the length of a walk $\omega$,
$x$ is the weight of each unit length,
and the summation is over all possible self-avoiding paths. 
In two and higher dimensions (D),
the SAW has two distinct phases separated by a critical point $x_c$.
The length $|\omega|$ remains finite in the dilute phase with $x <x_c$, 
and becomes divergent in the dense region with $x > x_c$.

Markov chain Monte Carlo (MCMC) methods have been extensively used in the simulation of the SAW \cite{Rensburg09}. 
The balance condition (BC) and the ergodicity are two key factors in designing a MCMC algorithm. 
The BC states that the probability flow entering into a configuration equals the flow out of the configuration. Thus, it ensures a stationary distribution. Then, ergodicity ensures convergence to the distribution \cite{Norris98}.  In practice, the BC is typically satisfied by employing the detailed balance condition (DBC), which implies that the probability flow from one configuration to another is equal to the reverse flow, i.e. the dynamics is reversible. 

In recent years, there have been several 
successful studies \cite{ST10, TS13, TCV11, FW11, BKW09, EAGIBK13, MKK14, KK15, MMK15, NMKH15} 
that show a promising future for MCMC algorithms beyond the DBC.
Geometric allocation approaches have been applied to the Potts model \cite{ST10,TS13};
irreversible MCMC methods have been designed for 
the mean-field Ising model \cite{TCV11,FW11}; event-chain Monte Carlo (ECMC) methods have been proposed 
for the simulation of hard-sphere systems \cite{BKW09,EAGIBK13} and generalized to particle systems with arbitrary pairwise interactions \cite{MKK14}, including soft-disk systems \cite{KK15}, 
the XY model \cite{MMK15} and the Heisenberg model \cite{NMKH15}.
Near the phase transition point, the geometric allocation method 
outperforms the standard Metropolis-Hastings (MH) method 
by $6.4$ times for the $q=4$ square lattice Potts model, 
and its performance increases with $q$ \cite{ST10, TS13}.
For the mean-field Ising model, the irreversible MCMC method has a dynamic exponent $z \simeq 0.85$,
which is considerably smaller than $z \simeq 1.43$ for the reversible MH method \cite{TCV11,FW11}. 
In comparison with the MH method, the speedup of the ECMC method reaches two orders of magnitude
in large systems consisting of $10^6$ hard spheres \cite{EAGIBK13}. 
For the 3D ferromagnetic Heisenberg model, it has been
reported that the ECMC method has a dynamic exponent $z \simeq 1$,
in contrast to $z \simeq 2$ for the MH method \cite{NMKH15}. 

For a Monte Carlo Markov chain, let $\pi (\omega)$ be the weight of a configuration $\omega$, 
$A(\omega \rightarrow \omega')$ be the a priori probability of proposing a transition from 
a configuration $\omega$ to another  $\omega'$,
and $P(\omega \rightarrow \omega')$ be the probability of accepting the proposal.
One obtains the stationary probability flow 
$\phi(\omega \rightarrow \omega') \equiv
\pi(\omega)A(\omega \rightarrow \omega') P(\omega \rightarrow \omega')$.
The DBC states that for any pair of $\omega$ and $\omega'$, 
$\phi(\omega \rightarrow \omega') =  \phi(\omega' \rightarrow \omega)$.
Instead, the BC requires that for any  $\omega$, 
$\sum_{\omega'} \phi(\omega \rightarrow \omega') 
= \sum_{\omega'} \phi(\omega' \rightarrow \omega)$,
with the summation over all possible configurations $\omega'$.  
Without the DBC, net probability flows can exist between two states $\omega$ and $\omega'$, 
i.e. $\phi(\omega \rightarrow \omega') \neq \phi(\omega' \rightarrow \omega) $, 
and the probability fluxes make circles in a phase space \cite{TCV11}.

While local probability flow circles are introduced in the geometric allocation approach,
considerably larger or even global circles can appear in other methods beyond the DBC. 
This is achieved through a lifting technique,  
 which enlarges the phase space by an auxiliary variable.
As a result, the system can be at different modes specified by different values of the auxiliary variable.
Within a given mode, an a priori direction of Monte Carlo updates is preferred,
and net probability flow may exist.
The BC is recovered by allowing switches between different modes, 
and the probability flow can satisfy a skew detailed balance (see e.g. Refs.~\onlinecite{SH13,HS13}).
For example,  the phase space  for the mean-field  Ising model \cite{TCV11}
is doubled and denoted as the decreasing and increasing modes.
In the first mode, only the positive spins are flipped and total magnetization $M$ decreases, 
whereas in the second mode,  $M$ increases by only allowing the flip of negative spins.
Updates can persist very far in one mode until a spin-flip proposal is rejected,
after which the updates are switched to the other mode.
This leads to large probability flow circles, 
and the diffusive feature of random updates is suppressed or even replaced by ballistic-like behavior.
It is noted that while the efficiency for the mean-field Ising model improves qualitatively, 
the lifting technique does not help significantly for the 2D Ising model  \cite{FW11,SB15}.

In this work, we design an irreversible MCMC algorithm for the SAW  by employing the lifting technique.
The update direction is selected such that the length $|\omega|$ increases
in one mode and decreases in the other. 
The two modes are ``linked" to each other through switching.
For practical coding,  only a few lines need be added to 
the widely-used  Berretti-Sokal (BS) algorithm, a simple variant of the MH method \cite{BS85}.
Nevertheless, numerical results show that the irreversible MCMC method 
is considerably superior to the BS algorithm. We use this new method to explore
the finite-size scaling (FSS) of the 5D SAW,  above the upper critical dimension $d_{\rm u}=4$. 
The critical point is located with high precision.

The rest of this article is organized as follows: In Sec.~\ref{sec_MH}, we review 
a few Monte Carlo methods for the SAW, including the conventional MH algorithm and the BS method. 
Section~\ref{sec_IL} describes the irreversible MCMC algorithm. 
We compare the performances of these algorithms in Sec.~\ref{sec_Per}. 
Section~\ref{sec_fss} contains a FSS analysis for the SAW on a periodic hypercubic lattice in five dimensions.
A brief conclusion and discussion are presented in Sec.~\ref{sec_con_dis}.

\section{Reversible MCMC algorithms}
\label{sec_MH}
We review below the MH and BS algorithms for the SAW, which employ the DBC
and are reversible.

 {\bf [Metropolis-Hastings algorithm.]} Given a regular $d$-dimensional lattice with coordination number $z$, 
a SAW $\omega$, with length $\calN \equiv |\omega|$, is a sequence of $\calN + 1$ lattice sites connected via a chain of occupied edges.
For simplicity, we fix an end of the walk at the origin and denote the movable end by $I$.
The MH algorithm \cite{Metropolis53,Hasting70} is  a standard reversible MCMC algorithm,
constructed as follows.


\noindent\rule{8.6cm}{0.5pt}
{\bf Metropolis-Hastings Algorithm}
\begin{enumerate}
\item Randomly select one of the $z$ neighboring sites of $I$, e.g., $I'$, 
and propose a symmetric update of the edge in between, which flips an empty edge to be occupied, and vice versa.
\item If the update leads to a valid SAW $\omega'$, accept the proposal with probability  $P={\rm min} \{1, \pi(\omega')/\pi(\omega) \}$.
\end{enumerate}
\noindent\rule[2mm]{8.6cm}{0.5pt}
The weight $\pi (\omega)$ of a configuration  is just given by Eq.~(\ref{partition_sum}) as $\pi(\omega) = x^{|\omega|}$,
and the acceptance probabilities are 
\begin{eqnarray}
P(\Delta \calN = +1) &=& \textrm{min} \{1, x \}  \nonumber \\
P(\Delta \calN = -1) &=& \textrm{min} \{1, 1/x \} \;,
\end{eqnarray}
where $\Delta \calN = |\omega'|-|\omega|$.
Here, the a priori probability is a constant $A(\omega \rightarrow \omega')= 1/z$, independent of  $\Delta \calN = \pm 1$. 
The DBC  can be easily proven.

{\bf [Berretti-Sokal algorithm.]} 
On a lattice with large coordination number $z$ (e.g., a high-dimensional lattice), 
the critical point of the SAW occurs  at $ x_c \approx 1/(z-1)$. 
This implies that in the above MH algorithm a typical update would attempt to flip an empty edge but fail, 
and is thus ineffective.
A more efficient MH  algorithm  for SAWs is the BS method \cite{BS85}, 
which employs a different scheme for a priori probabilities. 
A version of the BS algorithm, slightly different from that in Ref.~\onlinecite{BS85}, is described below.

\noindent\rule{8.6cm}{0.5pt}
{\bf Berretti-Sokal Algorithm} 
\begin{enumerate}
\item Choose Action: propose with equal probability the ``add" action $a_+$ for $\Delta \calN = +1$ and 
the ``delete" action $a_-$ for $\Delta \calN = -1$. 
\item Perform Action: 
\begin{itemize} 
 \item For action $a_+$,   randomly occupy with probability $P_{\rm BS}^+$ one of the $(z-1)$  empty edges if it leads a valid SAW. 
 \item  For action $a_-$,   delete with probability $P_{\rm BS}^-$ the last occupied edge, incident to $I$.
\end{itemize} 
\end{enumerate} 
\noindent\rule[2mm]{8.6cm}{0.5pt}
Note that an empty edge is chosen with probability $1/2(z-1)$ while 
the occupied edge is chosen with probability $1/2$.
The DBC condition leads to the Metropolis acceptance probabilities as
\begin{eqnarray}
P_{\rm BS}^+ &=& \textrm{min} \{1, x(z-1) \} \nonumber \\
P_{\rm BS}^- &=& \textrm{min} \{1, 1/x(z-1) \}  \;.
\label{eq-Pdec}
\end{eqnarray}
Near the critical point, the acceptance probabilities are close to unity.
In the original version \cite{BS85}, a priori probabilities 
are slightly further optimized such that the BS algorithm becomes rejection-free.

It is noted that some special attention is needed for the case of a ``null" walk with $\calN=0$, 
in which there are $z$ empty incident edges instead of $z-1$. 
For simplicity, we simply permanently delete an edge incident to the original site.

The diffusive feature of the BS algorithm is rather obvious: particularly for $x(z-1) \approx 1$ ,
 a SAW with length $\calN$ will have length $\calN' = \calN \pm 1$ in the next step
with approximately equal probabilities. 
 
\section{Irreversible MCMC method}
\label{sec_IL}
\begin{figure}
 \begin{center}
  \includegraphics[angle=-90,width=8.5cm]{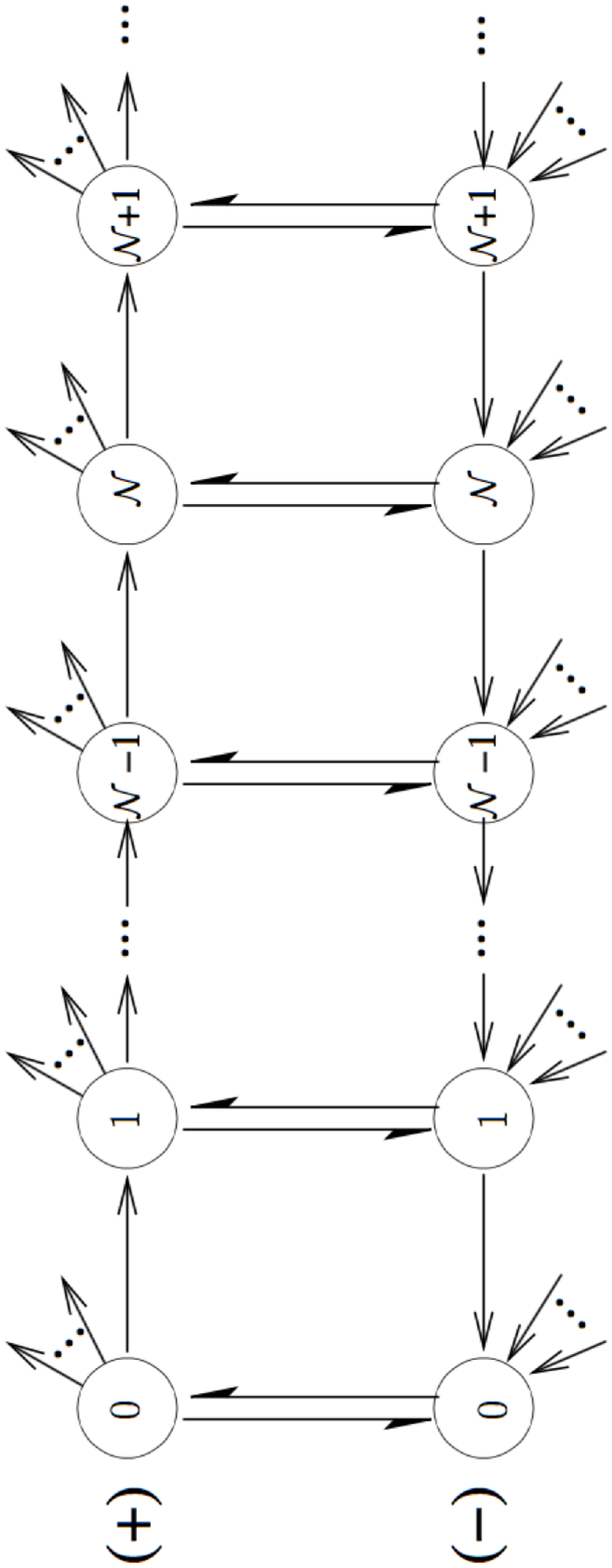} 
 \caption{Sketch of probability flows in the irreversible Monte Carlo algorithm for the SAW.
The enlarged state space consists of two modes labelled as $(+)$ and $(-)$, 
and each configuration is denoted by a circle with the inside number for the length $\calN$. 
 Action $a_-$, which decreases the length by a unity, is forbidden in  the increasing mode $(+)$, and  vice versa.}
 \label{fig_illu}
 \end{center}
\end{figure}

 A direct approach toward an irreversible method is to double the state space
of the SAW by introducing an auxiliary variable with two values $(+)$ and $(-)$.
In the increasing mode $(+)$, action $a_-$ is forbidden and the walk length $\calN$ increases 
because of action $a_+$. 
In contrast, $\calN$ decreases in the decreasing mode $(-)$. 
The balance condition is satisfied by allowing switching between the two modes.
Figure \ref{fig_illu} shows a sketch of the associated probability flows.
The formulation of the irreversible algorithm is considerably similar to the above BS method, 
as illustrated below.
\noindent\rule{8.6cm}{0.5pt}
{\bf Irreversible Algorithm}
\begin{itemize}
\item For the increasing mode $(+)$, perform action $a_+$ with probability $P^{+}$.
Randomly select and occupy one of the $(z-1)$ empty edges  
if this leads to a valid SAW; otherwise, switch to  the decreasing mode $(-)$. 
\item For the decreasing mode $(-)$, perform action $a_-$ with probability $P^{-}$. 
Delete the last occupied edge in the SAW if $\calN >0$; otherwise, switch to  the increasing mode $(+)$.
\end{itemize}
\noindent\rule[2mm]{8.6cm}{0.5pt}
The acceptance probabilities, $P^+$ and $P^-$, can also be given by Eq.~(\ref{eq-Pdec}). 
For the case where $x(z-1)<1$, the SAW grows with probability $P^+ = x(z-1)$ in the increasing mode $(+)$,
and it is deleted until $\calN=0$ in the decreasing mode $(-)$. 
For the case where $x(z-1) > 1$, which includes the critical point $x_c$, 
one obtains $P^+=1$. Thus in the increasing mode $(+)$, 
the SAW grows until it violates the self-avoidance, after which it switches to the decreasing mode $(-)$,
where the chain decreases with probability $P^- = 1/x (z-1)$.
At the critical point $x_c$, $P^-$ is close to but smaller than one, 
for example $P^- \simeq 0.9823$ in 5D. Thus, instead of returning to $\calN=0$ and growing again, 
it decreases for a considerably long time, 
then it switches to the increasing mode $(+)$ and grows again.
It is clear that the diffusive feature of Monte Carlo updates 
in the state space is significantly suppressed.

We demonstrate below the balance condition for the case where $x(z-1) <1 $, 
using the acceptance probabilities given by Eq.~(\ref{eq-Pdec});
the proof for $x(z-1) \ge 1$ follows the same  procedure.
Consider a $\calN$-step SAW  in the increasing mode $(+)$, 
the incoming probability flow from the $(\calN - 1)$-step SAW  is 
\begin{equation}
\phi^{(+)}_{{\rm in},a_+} = [1/(z-1)] x^{\calN-1} P^{+} = x^\calN \;.
\end{equation}
The factor $1/(z-1)$ accounts for the probability for selecting the particular edge leading to  the current SAW. 
The incoming probability flow due to the switch from mode $(-)$ is equal to zero unless $\calN=0$: 
$\phi^{(+)}_{\rm in, s}  = x^{\calN} (1-P^-) = 0 $.

The total outgoing probabilities flows are clearly $x^\calN$, since no action is allowed to keep the configuration unchanged. 
In the next step, there is an $(\calN+1)$-step SAW in mode $(+)$  or an $\calN$-step SAW in mode $(-)$.
More specifically, suppose that occupying one of $z' \in [0, z-1]$ empty edges would lead to a valid SAW, 
the outgoing probability flow is $\phi^{(+)}_{{\rm out}, a_+}=  [z'/(z-1)] x^\calN P^+ = z' x^{\calN+1} $ 
and the switch probability flow is $\phi^{(+)}_{\rm out, s} = x^\calN (1-z' x)$.
Thus, the balance condition in the increasing mode $(+)$ is obviously satisfied as  
\begin{equation}
\phi^{(+)}_{{\rm in}, a_+} + \phi^{(+)}_{\rm in,s} 
=  \phi^{(+)}_{{\rm out}, a_+} + \phi^{(+)}_{\rm out,s} = x^\calN \; .
\end{equation}
The same procedure follows for the BC in mode $(-)$.

 \begin{table}
 \caption{\label{tab:critical-points} 
The critical point $x_c$ for SAW on $d$-dimensional hypercubic lattices.} 
 \begin{tabular}{ll}
 \hline
 $d$ & $x_c$  \\
 \hline
 $2$ & $0.379\,052\,277\,758(4)$~\cite{Jensen03} \\
       &  $0.379\,052\,277\,755\,162(4)$~\cite{JSG16} \\
 $3$ & $0.213\,491\,0(3)$~\cite{HG04} \\
 $4$ & $0.147\,622\,3(1)$~\cite{OP01} \\
 $5$ & $0.113\,140\,81(4)$~\cite{OP01}\; \;\; $0.113\,140\,843(5)$~\textrm{[this work]}  \\
 \hline
 \end{tabular}
 \end{table}

\section{Performance}
\label{sec_Per}
We conducted simulations for the SAW at criticality on a $d$-dimensional periodic hypercubic lattice from $d=2$ to $5$,
where the critical value $x_c$ is listed in Table~\ref{tab:critical-points}.
For each linear size $L$, we carried out $5 \times 10^6/2^d$ sweeps ($L^d$ Monte Carlo steps) of simulations, 
in which one fifth were thrown for thermalization. 
The number of Monte Carlo steps between successive samples is $L/2$ for $d=2,3$, 
and $L^2/4$ for $d=4,5$. 

We compare the efficiency of the algorithms according to the {\it integrated} autocorrelation time $\tau$ for 
an arbitrary observable, defined by~\cite{MKB73}
\begin{equation}
\delta \calO = \sqrt{ \frac{1+ 2\tau/\Delta \tau }{n-1} ( \overline{\calO^2} - {\overline{\calO}}^2 ) } \; ,
\label{def_tau}
\end{equation}
where  $\delta \calO$ is the standard deviation,  $\Delta \tau$ denotes the number of sweeps between successive samples, 
and $n$ is the number of samples.
For $\Delta \tau \gg \tau$, i.e., the successive samples are effectively independent, Eq.~(\ref{def_tau}) 
is simplified as $\delta \calO = \sqrt{ ( \overline{\calO^2} - {\overline{\calO}}^2 ) / (n-1) }$.


In simulation, we sampled the walk length $\calN$ and observable $\calD_0$ that describes the event of a null SAW: $\calD_0=1$ for $\calN = 0$
and $\calD_0=0$ otherwise. The statistical average $D_0 = \langle \calD_0 \rangle$  accounts for the probability that the walk end $I$ returns to the original site. 

\begin{figure}
 \begin{center}
 \includegraphics[width=8.0cm]{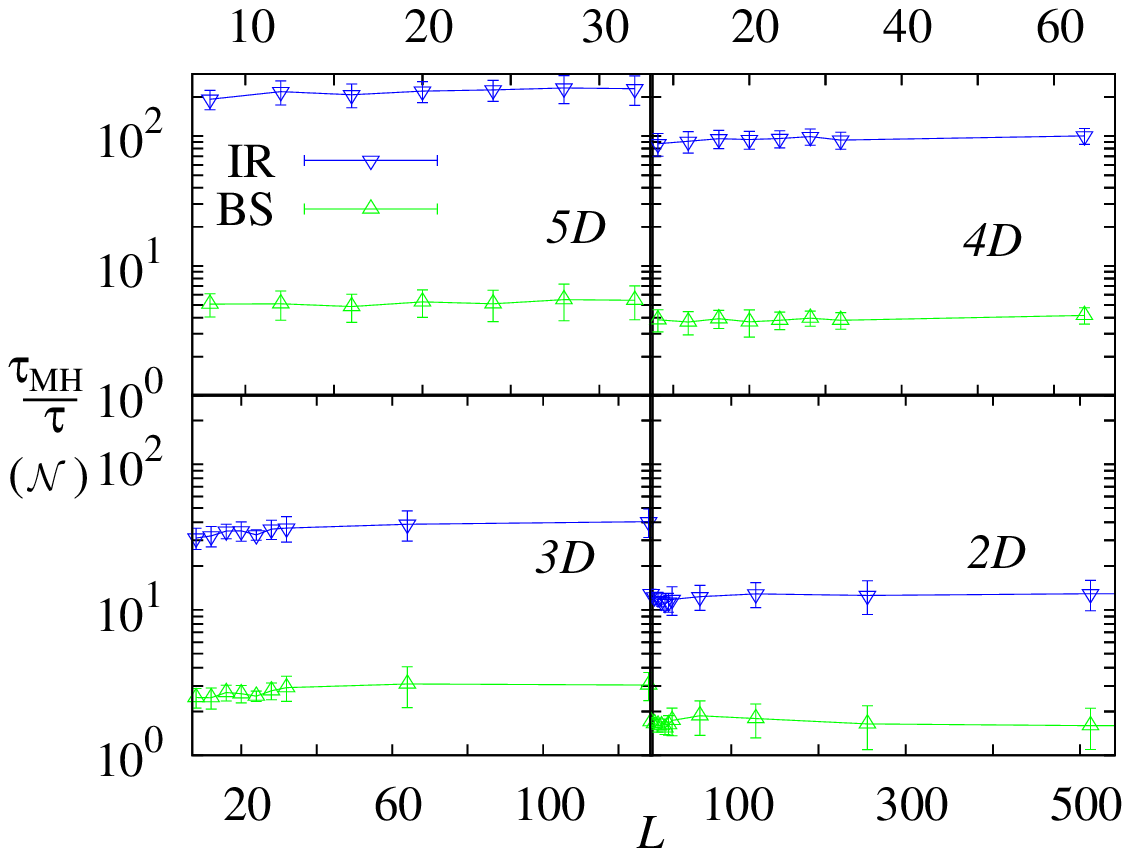} 
 \caption{Autocorrelation time  $\tau(\calN)$ for the MH algorithm $\tau_{\rm MH}$ divided by that of the BS or the irreversible (IR) algorithm, 
 versus the linear system size $L$.}
\label{fig_tau}
 \end{center}
\end{figure}

\begin{figure}
 \begin{center}
 \includegraphics[width=8.0cm]{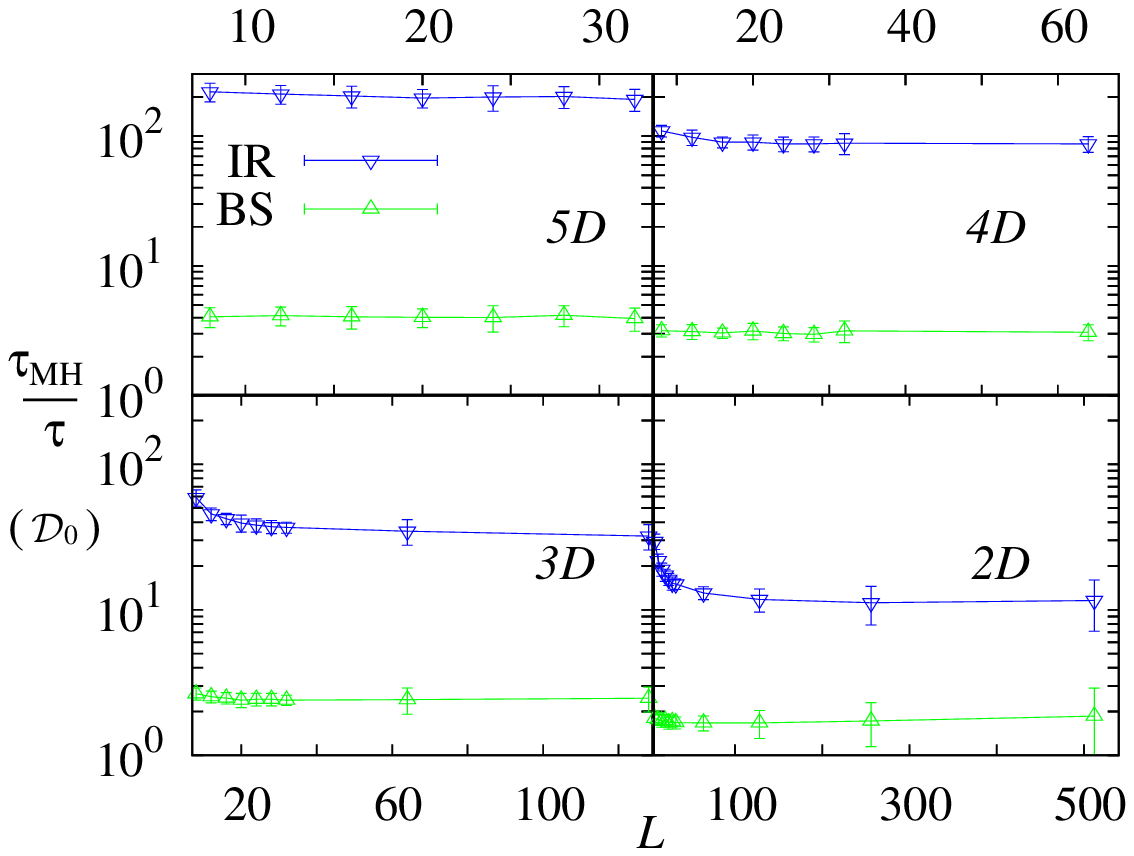} 
 \caption{Autocorrelation time $\tau_{\rm MH} (\calD_0) $ of the MH algorithm  divided by that of the BS or the irreversible (IR) algorithm, 
 versus the linear systems size $L$.}
 \label{fig_tau_r}
 \end{center}
\end{figure}

The autocorrelation time $\tau$ is measured for $\calN$ and $\calD_0$. 
Figure~\ref{fig_tau} and ~\ref{fig_tau_r} compare the autocorrelation times $\tau(\calN)$ and $\tau(\calD_0)$ for the MH, BS, and irreversible algorithm.
It can be seen that the performance of  the irreversible algorithm is considerably superior to the BS and the MH algorithm. 
The gained efficiency becomes more pronounced as $d$ increases. 
For $d=2$, it outperforms the MH and BS algorithm by approximately $12$ and $8$ times, respectively, 
while they become approximately $200$ and $40$ for $d=5$.  
As an illustration, Table~\ref{tab:tau_N} shows $\tau (\calN)$ for the various algorithms, for the maximum linear size $L_{\rm max}$ in this comparison study. 
We also compare the efficiency in terms of CPU time using a desktop computer with $3.8$ GiB memory and four Intel i5 cores.
As shown in the last column in Table~\ref{tab:tau_N},  the results are similar to those for the walk length $\calN$.

 \begin{table}
 \caption{\label{cmp_CPU_time} Autocorrelation time $\tau (\calN)$ for the various algorithms, 
 in unit of sweeps. For convenience, the ratio of CPU time $t_{\rm BS}/t_{\rm IR}$ is also calculated.}
 \label{tab:tau_N}
\begin{tabular}{lllllll}
\hline
 $d$ & $L$ & $\tau_{\rm IR}$ & $\tau_{\rm BS}$ & $\tau_{\rm MH}$ & $\tau_{\rm BS}/\tau_{\rm IR}$ & $t_{\rm BS}/t_{\rm IR}$ \\
\hline
 $2$ & $1024$ & $12.9(3)$ & $107(4)$ & $166(6)$ & $8(2)$ & $15(2)$  \\ 
 $3$ & $128$ & $0.559(6)$ & $7.41(7)$ & $22.5(9)$ & $13(1)$ & $18(2)$  \\
 $4$ & $64$ & $0.087\;1(8)$ & $2.10(2)$ & $8.7(2)$ & $24(2)$ &  $17(2)$  \\
 $5$ & $32$ & $0.028\;2(4)$ & $1.20(3)$ & $6.5(3)$ & $43(6)$ & $35(6) $  \\
\hline
 \end{tabular}
 \end{table}

\section{Finite-size scaling above the upper critical dimension}
\label{sec_fss}
The method of FSS, which is derived from the renormalization-group theory, plays a fundamental role in 
numerical study of critical phenomena. 
It predicts that near $x_c$, the energy-like quantity, the specific-heat-like quantity and the magnetic susceptibility, 
$E$, $C$ and $\chi$,  
scale as 
\begin{eqnarray} 
E(x, L) &=& L^d E_{\rm r} (x) + L^{1/\nu}       E_{\rm s} [ L^{1/\nu} (x-x_c)] \nonumber \\
C(x, L) &=& C_{\rm r} (x) + L^{\alpha/\nu} C_{\rm s} [ L^{1/\nu} (x-x_c)] \nonumber \\
\chi (x, L) &=& \chi_{\rm r} (x) + L^{\gamma/\nu} \chi_{\rm s} [ L^{1/\nu} (x-x_c)] \; ,
\label{eq:FSS_0}
\end{eqnarray}
where the critical exponents, $\alpha$ and $\gamma$, are for the thermodynamic (i.e., infinite system size)
 quantities $C$ and $\chi$, respectively.  
In Eq.~(\ref{eq:FSS_0}), the first term accounts for the regular functions that are size-independent (except that the regular part of the energy is proportional to $L^d$), 
while $E_{\rm s}$, $C_{\rm s}$ and $\chi_{\rm s}$ are universal functions accounting for the singular behavior. 

The FSS formula~(\ref{eq:FSS_0})  is correct in dimensions lower than the upper critical dimensionality $d_{\rm u}$.
For $d>d_{\rm u}$, the thermodynamic critical exponents take mean-field values, 
but the FSS behavior is much more complicated. 
Different FSS behaviors  occur for different boundary conditions, and for the ${\bf k}=0$ and the ${\bf k} \neq 0$ fluctuations.
Extensive studies have been carried out for the 5D Ising model--i.e., the $n=1$ case of the O($n$) model,
for which the renormalization-group theory gives $d_{\rm u}=4$,  and mean-field exponents $\nu=1/2$, $\gamma=1$ and $\alpha=0$.
For periodic boundary conditions, the FSS formula ~(\ref{eq:FSS_0}) holds if the critical exponent 
$\nu=1/2$ is replaced by a renormalized one $\nu^* = 2/d$, although there still exist some debates on the involved physical scenarios \cite{BNPP85, BKW12, FBKW16, WY14}. 

As the Ising model, the SAW is also a special case of the O($n$) model in the $n \rightarrow 0$ limit. 
Using the irreversible algorithm, we performed extensive simulations for the SAW 
on a 5D periodic hypercubic lattice up to $L = 128$.
This provides an independent and accurate study of the $d>d_{\rm u}$ FSS behavior for the O($n$) universality.
In comparison with the Ising model, such a study of the 5D SAW has a few advantages.
The regular terms $C_{\rm r} (x)$ and $\chi_{\rm r} (x)$ in Eq.~(\ref{eq:FSS_0}) are supposed to vanish for $x \leq x_c$, 
since the average walk length per site $N/L^d = \langle \calN \rangle /L^d$ approaches to zero as $L$ increases ($N$ is 
the energy-like quantity with a zero regular part).
Further, the simulation can reach a relatively large linear size, while it is mostly restricted to $ L < 40$ for the 5D Ising model.

\begin{figure}
 \begin{center}
 \includegraphics[width=8.0cm]{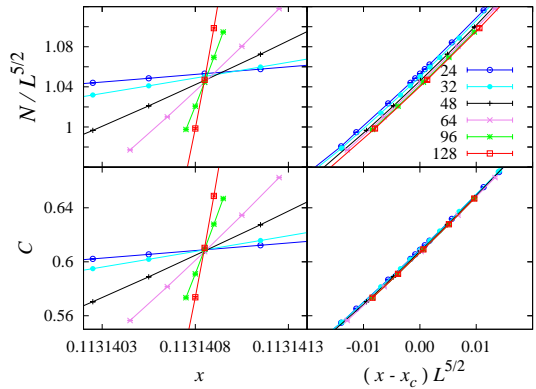} 
 \caption{Walk length $N / L^{5/2}$ (top) and the specific heat $C$ (bottom) for the 5D SAW.
 The right panels use $(x-x_c)L^{5/2}$ as the horizontal scale so that the data for 
 different $L$ collapse approximately into a single curve.}
 \label{fig_fss_5D}
 \end{center}
\end{figure}
\begin{figure}
 \begin{center}
 \includegraphics[width=8.0cm]{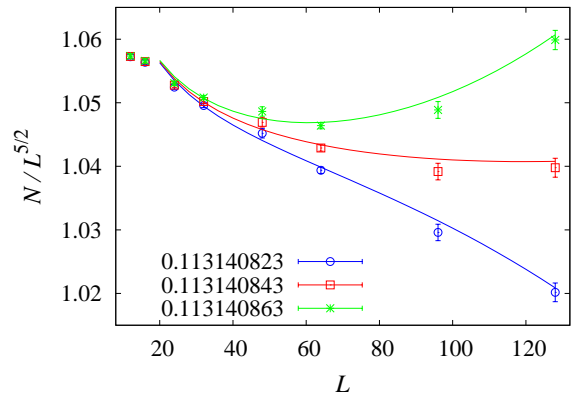} 
 \caption{Walk length $N/L^{5/2}$ versus $L$ for the 5D SAW.
The curves correspond to fitting results of the Monte Carlo data.}
 \label{fig_xc_demon}
 \end{center}
\end{figure}

We sampled the average walk length $N$, the specific-heat-like quantity $C = L^{-d} \left( \langle \calN^2 \rangle - \langle \calN \rangle ^2 \right)$
and the returning probability $D_0 = \langle \calD_0 \rangle$, which is the inverse of the magnetic  susceptibility $D_0 = 1/\chi$\;\cite{DGS07}.
Taylor expansion of Eq.~(\ref{eq:FSS_0}) leads to 
\begin{eqnarray}
N &=& L^{1/\nu^*} [ n_0 + \sum_{k=1}^{2} n_k (x-x_c)^k L^{k/\nu^*} + b_n L^{y_1} ]  \nonumber \\
C &=& c_0 + \sum_{k=1}^{2} c_k (x-x_c)^k L^{k/\nu^*} + b_c L^{y_1} \; ,
\label{eq:fit0}
\end{eqnarray}
where the term with exponent $y_1$ accounts for finite-size corrections and 
the coefficients of each term are non-universal. 
Besides the renormalized exponent $1/\nu^*=d/2$,
the leading correction exponent is predicted as $y_1= (d_{\rm u}-d)/2=-1/2$ ~\cite{BZ85, WY14}.
The FSS behavior of $C$ in Eq.~(\ref{eq:fit0}) is based on the prediction $\alpha/\nu^*=0$.
The data of $N$ and $C$ are shown in Fig.~\ref{fig_fss_5D}, in which
 $x_c$ is located by the approximate intersection point for different sizes,
and the rescaled plots in the right panels imply the correctness of Eq.~(\ref{eq:fit0}). 
The MC data were fitted by Eq.(\ref{eq:fit0}) according to the least-squared criterion,
and the results are shown in Table \ref{tab:fit_results}. 
The correction exponent $y_1$ is fixed at the predicted value $-1/2$ in the fits. 
As a precaution against correction-to-scaling terms not included in the fit ansatz, 
we imposed a lower cut off $L \ge L_{\rm min}$ on the data points, 
and observed the change of $\chi^2/DF$, where ``DF" represents the number of degrees of freedom.
The estimates of $1/\nu^*$ agree well with the predicted value $5/2$.
We take the final determination  $x_c = 0.113\,140 \,843(5)$, 
which is significantly improved compared to the best available result  $x_c=0.113\,140\,81(4)$ \cite{OP01}. 
The reliability of our estimate of $x_c$ is further demonstrated in Fig.~\ref{fig_xc_demon}, 
which clearly shows that $x=0.113\;140\;823$ and $0.113\;140\;863$ are below and above the critical point, respectively.

 \begin{table*}[t]
 \centering
 \caption{Fit results for the walk length $N$ and specific heat $C$ of the 5D SAW.}
\label{tab:fit_results} 
\begin{tabular}{lllllllll}
\hline
& $x_c$ & $1/\nu^{*}$ & $n_0$ & $n_1$ & $n_2$ & $b_1$ & $L_{\rm min}$ & $\chi^2/DF$ \\ 
$N$ & $0.113\;140\;840 (2)$ & $2.500 (6)$ & $1.03 (4)$ & $5.4 (3)$ & $16(7)$ & $0.13 (8)$ & $32$ & $4/11$ \\
& $0.113\;140\;840 (2)$ & $5/2$ & $1.025 \;6 (10)$ & $5.37 (2)$ & $15(6)$ & $0.138 (6)$ & $32$ & $4/12$ \\
\hline
& $x_c$ & $1/\nu^{*}$ & $c_0$ & $c_1$ & $c_2$ & $b_c$ & $L_{\rm min}$ & $\chi^2/DF$ \\ 
$C$ & $0.113\;140\;843 (2)$ & $2.50 (2)$ & $0.602\;6 (11)$ & $4.0 (3)$ & $9(3)$ & $0.034 (6)$ & $32$ & $12/13$ \\
& $0.113\;140\;843 (2)$ & $5/2$ & $0.602\;6 (11)$ & $4.03 (2)$ & $10(2)$ & $0.034 (6)$ & $32$ & $12/14$ \\
\hline
 \end{tabular}
 \end{table*}

\begin{figure}
 \begin{center}
 \includegraphics[width=8.0cm]{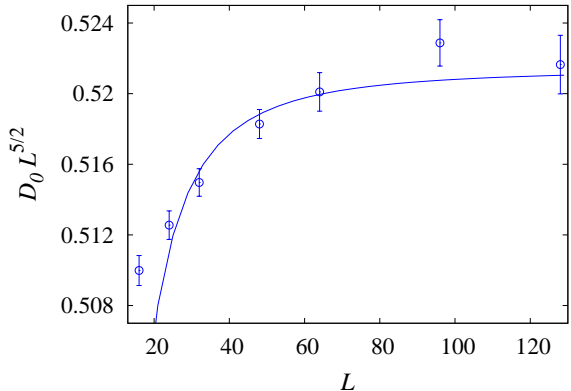} \\
 \caption{Returning probability $D_0 L^{5/2}$ versus $L$ at $x_c$ for the 5D SAW. 
The curve corresponds to fitting results of the Monte Carlo data.} 
 \label{fig-fss-5D-R}
 \end{center}
\end{figure}

At the critical point $x_c$, the FSS behavior of the returning probability $D_0$  should behave as
\begin{equation}
D_0  = L^{-\gamma/\nu^*} ( d_0 + d_1 L^{y_1} ) \;,
\end{equation}
with the mean-field value $\gamma=1$. 
This is confirmed by Fig.~\ref{fig-fss-5D-R}.

\section{Conclusion and discussion}
\label{sec_con_dis}
We develop an irreversible MCMC algorithm for the SAW.
It violates the DBC, and satisfies the weaker BC.
In comparison with the standard MH algorithm and one of its variants, i.e., the BS algorithm, 
the irreversible method is considerably more efficient. 
While the BS algorithm is approximately $d$ times more efficient than the standard MH algorithm,
the irreversible algorithm is further superior to the BS algorithm. 
The higher is the spatial dimension, the more is the gain of efficiency.
This is because the critical SAW is more like the ordinary random walk in higher dimensions, 
and thus the diffusive feature is more suppressed in the irreversible algorithm.

Using the irreversible MCMC algorithm for the SAW, we perform an independent and accurate test
of the renormalized exponents $1/\nu^*=d/2$ and $\gamma/\nu^*=d/2$ in the FSS behavior 
of systems in the O($n)$ universality class with periodic boundary conditions. 
We also provide an estimate $x_c =0.113\,140 \,843(5)$ for the 5D SAW, which is $8$ times more
precise than the best available estimate.

We believe that the irreversible MCMC algorithm for the SAW will make an important contribution 
toward a deeper understanding of the FSS behavior above the upper critical dimension, 
particularly for systems with free boundary conditions.
This irreversible MCMC algorithm must also be very valuable in studying interacting SAW \cite{Rensburg09}, polymeric systems and other soft matters.
Similar irreversible tricks can be introduced into other algorithms like the worm algorithm \cite{DGS07,worm1, worm2}.
These research activities are undergoing.

Several efficient Monte Carlo methods exist for simulating SAWs \cite{Rensburg09,Nidras96,MS88,Grassberger97}. 
In particular, the Pivot algorithm~\cite{MS88} and the pruned enriched Rosenbluth method (PERM) ~\cite{Grassberger97} 
are  known to be very efficient in the canonical ensemble where the SAW chain has a fixed length.
For example, the critical point $x_c$ for $d=3, 4, 5$ in Table~\ref{tab:critical-points} was obtained by PERM  \cite{HG04, OP01}. 
It would be interesting to study whether it is possible to implement the irreversible technique in these state-of-the-art algorithms for SAWs.

\begin{acknowledgments}
This work is supported by the National Natural Science Foundation of China 
under Grant No.~11275185 and No.~11625522, and by the Open Project Program of State Key Laboratory 
of Theoretical Physics, Institute of Theoretical Physics, Chinese Academy of 
Sciences, China (No.~Y5KF191CJ1).  
Y. Deng acknowledges the Ministry of Education (China) for 
the Fundamental Research Funds for the Central Universities under Grant No.~2340000034.
\end{acknowledgments}


\end{document}